\documentclass[referee]{raa}           
\usepackage{graphicx,times}
\usepackage{natbib}
\usepackage{amssymb,amsmath}
\bibpunct{(}{)}{;}{a}{}{,}

\usepackage[a4paper=true,dvipdfm=true,pagebackref=true]{hyperref}
\hypersetup{pdftitle = The title of my PDF, pdfauthor = My name, pdfsubject= The subject, pdfkeywords = keyword1 keyword2 keyword3} 
\hypersetup{colorlinks = true, linkcolor = green, anchorcolor = red, citecolor = blue, filecolor = red, pagecolor = red, urlcolor = red}

\begin{document}

   \title{A report on the type II X-ray burst from SMC X-1
%\footnotetext{\small $*$ Supported by the National Natural Science Foundation of China.}
}

 %\volnopage{ {\bf 20XX} Vol.\ {\bf X} No. {\bf XX}, 000--000}
   \setcounter{page}{1}

   \author{Binay Rai\inst{1}, Pragati Pradhan\inst{2,3}, Bikash Chandra Paul\inst{1}
   }

{   \institute{ Dept. of Physics, North Bengal University, Darjeeling, West Bengal, 734013, India; {\it binayrai21@gmail.com}\\
             \and
                Department of Astronomy and Astrophysics, Pennsylvania State University,  Pennsylvania, 16802, US\\
            \and
            St. Joseph's College, Darjeeling-734104, West Bengal, India\\
   \vs \no
%   {\small Received 2012 June 12; accepted 2012 July 27}
}

\abstract{We study RXTE PCA data for the high mass X-ray binary source SMC X-1 between 2003-10 and 2003-12 when the source was in high states. The source is found to be frequently bursting which can be seen as flares in lightcurves on an average of one in every 800 s, with an average of 4-5 X-ray burst per hour of type II. We note that typically burst was short lasting for few tens of seconds in addition few long bursts of more than hundred seconds were also observed. The flares apparently occupied 2.5$\%$ of the total observing time of 225.5 ks. We note a total of 272 flares with mean FWHM of the flare $\backsim$21 s. The rms variability and the aperiodic variability are independent of flares. As observed the pulse profiles of the lightcurves do not change its shape implying that there is no change in the geometry of accretion disk due to burst. The hardness ratio and the rms variability of lightcurves show no correlation with the flares. The flare-fraction shows a positive correlation with the peak-to-peak ratio of the primary and secondary peaks of the pulse profile. The observed hardening or the softening of the spectrum cannot be correlated with the flaring rate but may be due to the interstellar absorption of X-rays as evident from the change in the hydrogen column density ($n_{H}$). It is found that the luminosity of the source increases with the flaring rate. Considering the viscous timescale equal to
 mean recurrence time of flares we fixed the viscosity parameter $\alpha$ $\backsim$ 0.16.
}

   \authorrunning{Binay Rai et.al. }            %author_head in even pages
   \titlerunning{Type II X-ray burst from SMC X-1}  % title_head in odd pages
   \maketitle

%________________________________________________ sections below
%
\section{Introduction}           %% first-level sections will be auto-capitalized
\label{sect:intro}

$\;\;$ The SMC X-1  is one of the high mass X-ray binary (HMXB) system in SMC. The Type II X-ray burst from the source was discovered by Angelini {\it et al.} (1991) along with aperiodic variability of 0.01 Hz. It was discovered while compared the burst with MXB 1730--355 known as the ``Rapid Burster" (\citealt{Lewin+etal+1976}). The burst is considered to be due to a viscous instability in the accretion disk characterized by sharp rise and decay of count rate in the lightcurve with the recurrence time ranging from $\sim$ 10 s to 1 hr. 

$\:\:$ It may be pointed out here that Type I X-ray burst (\citealt{Lewin+etal+1983}) corresponds to thermonuclear origin with a sharp rise and exponential decay in the intensity observed in the lightcurve along with a recurrence time of hours or days. The spectrum of the Type I burst is consistent with that of a blackbody followed by spectral softening in burst decay. It was observed that in ``Rapid Burster" the average fluxes emitted in Type II X-ray bursts was 120 times the average fluxes emitted in Type I X-ray bursts 
(\citealt{Hoffman+etal+1978}). The observed value of the ratio of the time-averaged persistent flux to the time-averaged Type I X-ray burst flux lies between $10$ to $10^{3}$ (\citealt{Lewin+etal+1993}).

$\;\;$The SMC X-1 is  a binary system (HMXB) with a neutron star (\citealt{Price+etal+1971}) and B0 supergiant SK 160 of mass $\sim$17.2M$_{\odot}$ having an orbital period of $\sim$3.9 days (\citealt{Schreier+etal+1972}). The X-ray source is eclipsed by the companion for 0.6 days. The source is found to have a regular spin up state $\sim$3.279$\times$10$^{-11}  $Hzs$^{-1}$ (\citealt{Davison+etal+1977, Wojdowski+etal+1998}) with no spin down state as has been observed in the source. The source is found to have orbital decay rate of $\sim$3.4$\times$10$^{-6}$ yrs$^{-1}$                                  (\citealt{Levine+etal+1993, Wojdowski+etal+1998}). It has been reported that the SMC X-1 shows an aperiodic variation of $\sim$(55-60) days due to obstruction of the X-rays coming from the source by its tilted precessing accretion disk (\citealt{Wojdowski+etal+1998}). It is also regarded as an ``intermediate-stage source" (\citealt{Moon+etal+2003}) between low-mass X-ray binaries (LMXBs) and X-ray pulsars along with some interesting intermediate-stage sources  having magnetic field between 10$^{8}$ to 10$^{11}$ G and which undergoes either Type I or II X-ray burst or both and may or may not have coherent pulsation. The examples of these type of sources are ``the Bursting Pulsar" i.e. GRO J1744--28 (\citealt{Fishman+etal+1995}) which shows Type II character burst along with the coherent pulsation, ``the Rapid Burster" i.e. MXB 1730--355 
(\citealt{Lewin+etal+1976}) undergoing both Type I and Type II X-ray burst and ``the Accreting Milli-second pulsar" like SAX J1808.4--3658 (in \citealt{'t Zand+etal+2006}) which undergoes Type I X-ray burst with a coherent pulsation.

$\;\;$The SMC X-1 and GRO J1744--28  have some properties in common namely both of them undergoes Type II burst with coherent pulsation (\citealt{Li+etal+1997}). Their spin periods are also very small, for SMC X-1 it is $\backsim$0.71 s and that for GRO J1744-28 is $\backsim$0.47 s. They have steady spin up in their spin with the measured magnetic field value $\backsim$10$^{11}$G (\citealt{Bildsten+etal+1997}) and luminosity below an Eddington limit. However in the case of GRO 1744--28 burst occurs at higher rate than SMC X-1, although they differ in burst pattern too. The analysis of RXTE PCA data of the SMC X-1 by \citealt{Moon+etal+2003} found that flare occupies 3$\%$ of the total observing time and spreading to orbital phases and strongly correlated with the variability of the lightcurve. The flare fraction is found to increase with the peak to peak ratio of the pulse profile. The properties of the SMC X-1 does not differ very much during flaring period from that of the normal state. We studied here in  detail properties of the Type II X-ray burst from SMC X-1 making use of RXTE PCA data  between 2003-10 and 2003-12. The organization of the paper is as follows: In Section 2 we discuss the data reduction to analyze the source, in Section 3 the study of lightcurves with flares are presented. The Section 4 is concerned with the study of pulse profile, hardness ratio and spectrum of the source. Section 5 deals with the study of correlation of burst with different parameters and also studied the relation of luminosity with flaring rate and finally discussion is given in Section 6.

\section{Data Selection and Reduction}
\label{sect:Data sel}

\ 
$\;\;$To investigate SMC X-1 source we have used RXTE PCA data. The PCA has an arrays of the five proportional counter unit (PCU) namely PCU 0, PCU 1, PCU 2, PCU 3, PCU 4 (\citealt{Jahoda+etal+2006}). The PCU 0 suffered propane loss in the year 2000. PCU 3 and PCU 4 were regularly given rest to avoided breakdown. Out of above mentioned counter units data from PCU 2 will be employed here for analysis as it was only PCU unit that didn't suffer any breakdown and was operating all the time. The PCU unit is best calibrated counter unit as verified by fitting the crab spectrum simply with the power-law (PL). 

The data reduction were done using HEASOFT ver 6.11 . For spectral analysis standard 2 mode of RXTE PCA data were used  which  has 129 channels and default binning of 16 s , the data only from the top Xe-layer of the PCU 2 were considered. The response matrix for the top layer has been obtained using \emph{FTOOL PCARSP}. The background spectra were obtained using a tool \emph{RUNPCABACKEST} taking bright source background model. The source spectra were then obtained by subtracting these background spectra with the total spectra in \emph{XSPEC}. Here the energy range for a spectrum under consideration lies in the range 3-18 keV, we have not consider data above 18 keV as the spectrum in that case is dominated by background and good fitting may not be possible. A systematic error of 2$\%$ were added to all spectra.

$\;\;$The timing analysis are carried out using \emph{GOODXENON}  mode of RXTE PCA data. The lightcurves were extracted from the \emph{GOODXENON} data using the mission specific tool \emph{SEEXTRCT} for all available whole energy range and all layers of the PCU2 with background correction. The background correction has been made by subtracting the background lightcurve from the total lightcurve using \emph{lcmath}. The background lightcurve were extracted using the background model for the bright source. The reference frame of photon arrival time were transformed to barycentre with the help of \emph{FTOOL FAXBARY} using JPL DE400 ephemeries. The set of data where the elevation of telescope was $<$10 degree and 30 minutes after the passage from South Atlantic Anamoly (SAA) has been considered in this paper for data reduction and analysis. 

\section{Lightcurves and flares }

$\;\;$For our analysis along with flares searching all the RXTE PCA data available for SMC X-1 between 2003-10 and 2003-12 are taken. The data we have taken  lies in two high states h1, h2 as observed in the RXTE ASM lightcurve of the source as shown in Fig.1. The technique used by \citealt{Moon+etal+2003} for searching flares is followed in this work. All the photon counts which were above 3$\sigma$ level above the mean is considered to be flare in a lightcurve. The data were divided into 110 data segments, with each segment of duration $\backsim$2050 s. So we have 110 lightcurve each of them have been plotted and analyzed. 
\begin{figure}
\includegraphics[height=6cm,width=12cm]{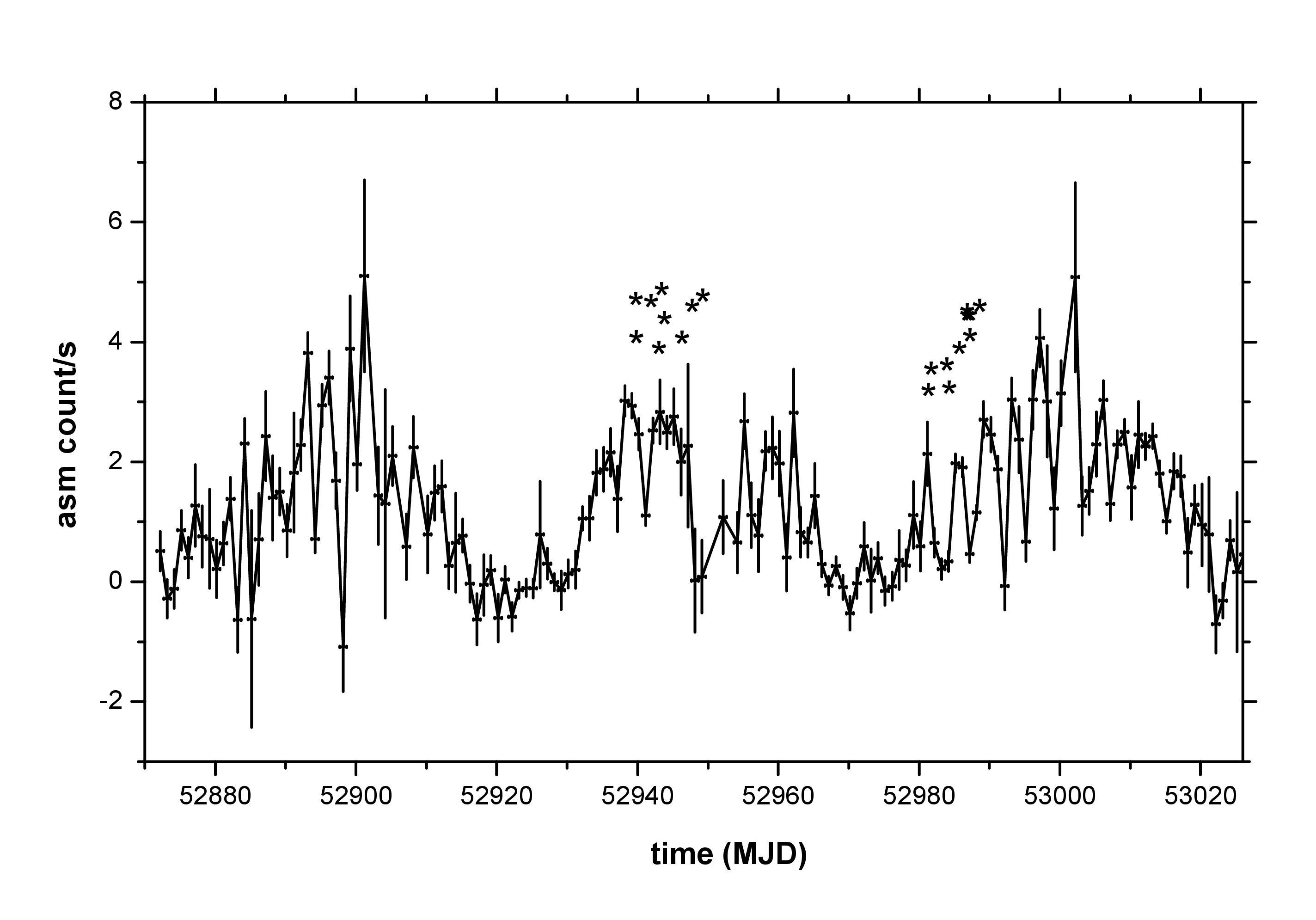}
\centering
\caption{\footnotesize A segment of RXTE ASM lightcurve of SMC X-1 with rebinning of 1 day, the ``*'' indicates the date of the observations. Our observation of the source fall in two high states indicated by h1 and h2.}
\end{figure}

$\;\;$The time spent by a burst varies from burst to burst where some of them last for few tens of seconds to more than that. To determine the duration of a burst we fit each flare with Gaussian model to obtain the width `$w$' of the flare. Consequently the FWHM of flare may be obtained by using the relation FWHM=$2.35482w$, which we take as a duration of a burst. The reduced $\chi^{2}$ of the fitting varies from 2.3 to 3.56. We began with the lightcurves of 4 s time resolution . We found 272 such cases of flares with mean FWHM of $\backsim$21 s and mean standard deviation of $\backsim$8 s. Therefore out of 225.5 ks we note 5.7 ks as the time of flaring. Thus its clear that the source is flare active for $\sim$2.5$\%$ of the total time which is close to the value reported earlier by \citealt{Moon+etal+2003}. The left of Fig.2 shows the lightcurves for four different observations with time resolution of 4 s and the corresponding normalised power spectra on the right. It is observed from the plot that the recurrence time between flares varies from few hundred to few thousand. The average value of recurrence time between the flares from the above analysis is $\sim$ 800 s. The number of flares, width of flares and recurrence time are found to be same for the lightcurves having 2 s, 6 s and 8 s time resolutions.

\begin{figure}
\includegraphics[height=9cm,width=9cm]{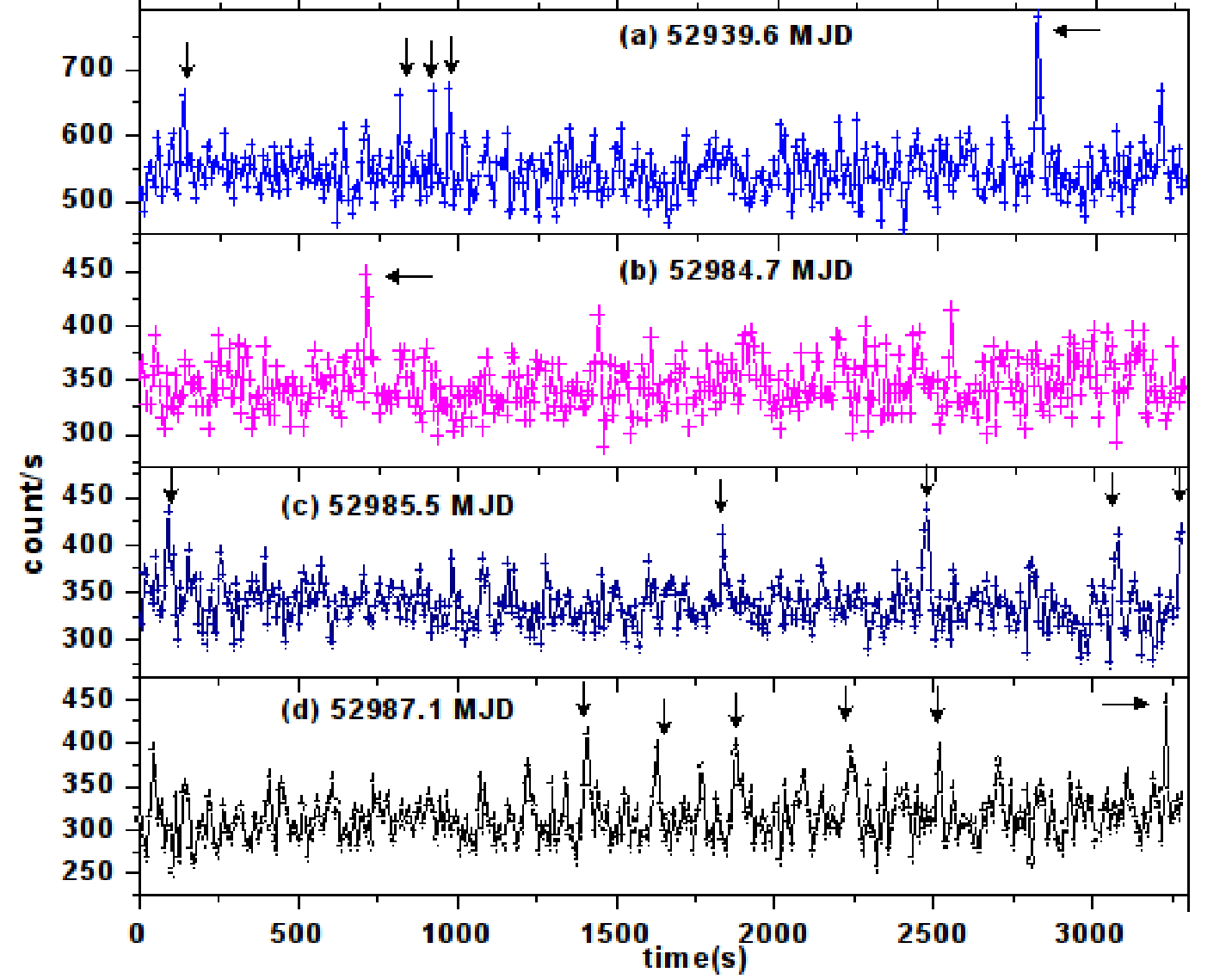}
\includegraphics[height=9cm,width=7cm]{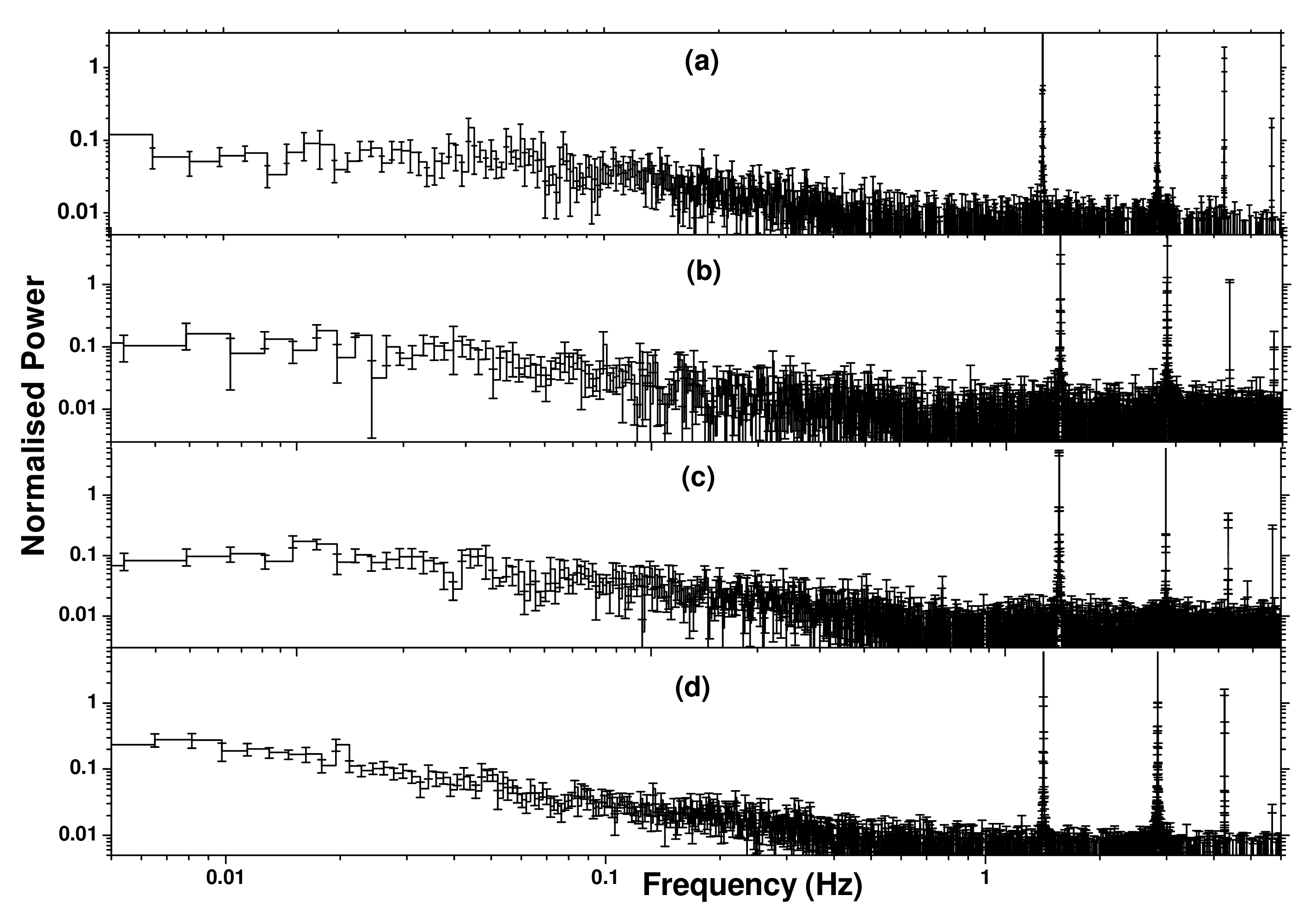}
\caption{\footnotesize The left figure is the lightcurves of four different observation indicated by lower case letters (a),(b),(c),(d) on the top of right hand sight along with date in MJD. The figure on the right  shows their corresponding  power spectra.The flares are indicated by an arrow in the lightcurves. The power in power density spectra are expressed in the unit of (rms)$^{2}$/Hz.}
\end{figure}
$\;\;$For power spectra we use data with the time resolution of 0.075 s and we plot them using XRONOS tool $\emph{POWSPEC}$ with normalization=-2. The value of normalization gives us the white noise normalised power spectra with their integral gives us square fractional rms variability. The power spectra corresponding to four lightcurves (left Fig.2) are shown to the right of fig.2. The rms variability of the four lightcurves obtained from the power spectra are estimated to be as (a)30.86$\%$,(b)31.2$\%$,(c)31.2$\%$ and (d)30.6$\%$. Thus during all four observations the source is found to show same variability i.e. the source were equally variable. Along with the coherent peak (at $\backsim$ 1.41 Hz) and its harmonic component evident from sharp rise of power in the power spectra along with some possible aperiodic components at low frequencies. For the power spectrum (d) it is noted that as we go to low frequency region (from 1 to 0.005 Hz) power rises moderately and at $\backsim$0.019Hz a broad peak exists. Similar rise is seen in right of figure 2(c) at the same frequency with flat top.

$\;\;$ It is observed that during some flares the count rate reached up to many times above the mean value, while in some cases flares were found to last for more than hundreds of second (Figure 3). Generally the bursts are single peaked (Figure 2) however large bursts were found to have multiple peaks before they finally decay to the mean value as shown in figure 3. The left side of Fig.3 shows a lightcurve with a flare of $\backsim$50 s consisting of two sharp peaks. The first peak arises because of sharp increase in the count rate reaching up to $\backsim$1350 s$^{-1}$ then sharply falls to  $\backsim$510 s$^{-1}$, followed once again by increase in count rate giving second peak of $\backsim$1320 s$^{-1}$ which falls rapidly below 500 s$^{-1}$. The situation is like one burst is followed by another one in no time. However the flare shown in the right side of the figure 3 is long $\backsim$300 s and has multiple peaks, the highest peak has a count rate of $\backsim$1000 s$^{-1}$ signalling the instability lasting for long time and  comes to normal stable state with slow rise and decay in intensity with multiple peak. The burst per hour per observation for the first high state (h1) as shown in Fig.1 is $\backsim$5 hr$^{-1}$ and for second (h2) we note $\backsim$4 hr$^{-1}$. However for total observation the average burst per hour was $\backsim$5 per hour and the average time between two burst may be set as $\backsim$800 s.
\begin{figure}
\centering
\includegraphics[height=6cm,width=10cm]{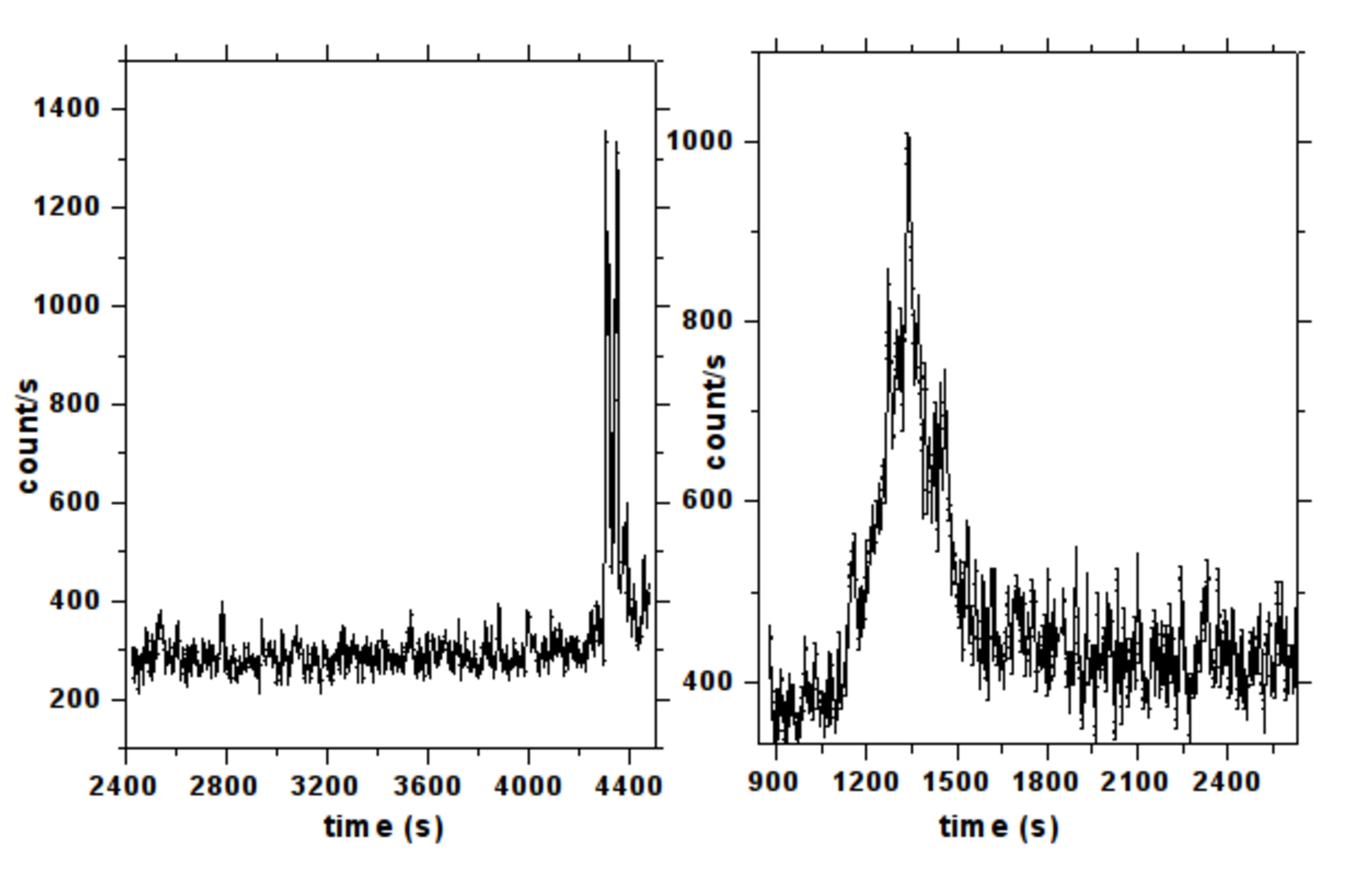}
\caption{\footnotesize Two large burst observed in SMC X-1. The Obs.Id for the observation is 80078-01-01-04. The left figure shows the burst where the count rate rises up to $\backsim$4 times above the mean value and in the right one the burst last for $\backsim$300 s. Both of the bursts have multiple peaks.}
\end{figure}
\section{Pulse profile, Hardness ratio and the energy spectra}

The binary orbit of the SMC X-1 is found to be nearly circular (\citealt{Levine+etal+1993, Wojdowski+etal+1998, Raichur+etal+2010}). As the neutron star moves in its circular binary orbit there are delays in the pulse arrival time because of orbital modulation \textit{i.e.} when the neutron star is close to an observer, the pulse will arrive sooner than when the neutron star is away from the observer. In order to get correct pulse profile the arrival time of pulse must be corrected so that we get an arrival time which also include an effect of orbital modulation.
If $t^{\prime}_{n}$ and $t_{n}$ be the time of emission and arrival respectively then they are related with each other and with the orbit of the neutron star through $f_{orb}(t^{\prime}_{n})$ (\citealt{Deeter+etal+1981}) as
\begin{equation*}
t^{\prime}_{n}=t_{0}+nP_{s}+\dfrac{1}{2}n^{2}\dot{P_{s}}P_{s}
\end{equation*}
\begin{equation*}
t_{n}=t^{\prime}_{n}+f_{orb}(t^{\prime}_{n})
\end{equation*}
\\
where $ P_{s}$ and $\dot{P_{s}}$ are respectively the spin period and the time derivative of the spin period of the neutron star, for a circular orbit $f_{orb}(t^{\prime}_{n})$ is of the form
\begin{equation*}
f_{orb}=a_{x} sini \ cos\l_{n}
\end{equation*}
\begin{equation*}
l_{n}=2 \pi (t_{n}^{\prime}-E)/P_{orb}+\pi/2
\end{equation*}
\begin{figure}
\centering
\includegraphics[height=5cm,width=10cm]{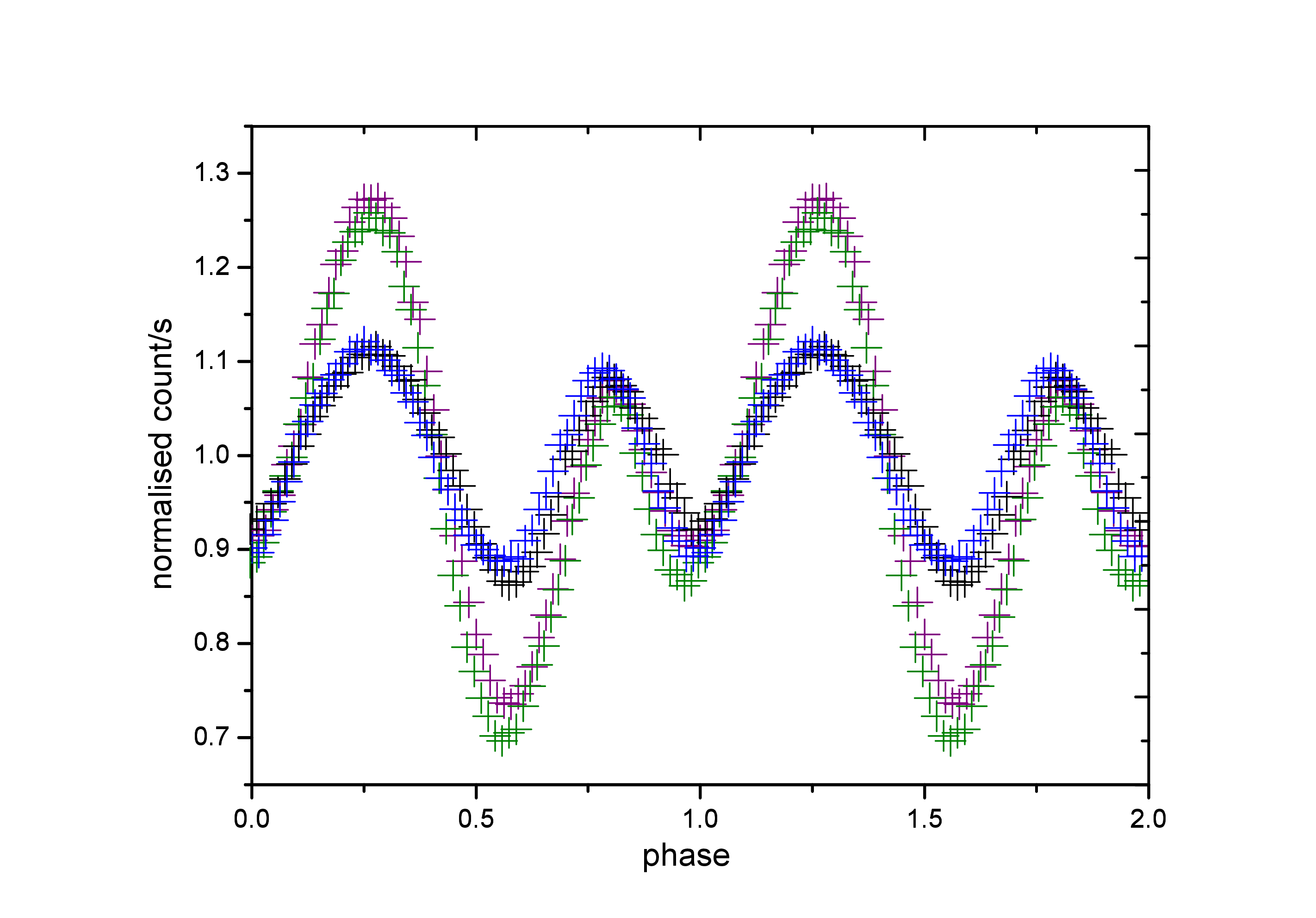}
\caption{\footnotesize Orbital corrected pulse profile of the four lightcurve of left of Fig.2(a),(b),(c) and (d) folded at $\backsim$0.7 s, blue (a),purple (b),green(c) and black (d)}
\end{figure} 

where $l_{n}$ is the mean orbital longitude at time $t_{n}^{\prime}$ and $E$ is the epoch when the mean orbital longitude is equal to $\pi/2$, $a_{x}$sin$i$ is the projected semi-major axis with `$i$' as an angle of inclination between the line of sight and orbital angular momentum vector. 
$\;\;$We have used the value of epochs and other orbital parameters from \citealt{Raichur+etal+2010} for the orbital correction. The orbital corrected pulse profiles are shown in figure 4. The pulse fraction of (a) and (d) is 20$\%$ where as for (b) and (c) its 30$\%$. All the pulse profiles have their secondary peak at a phase $\backsim$0.78 and primary at $\backsim$0.28. The secondary peaks of all four lightcurves coincides with each others. Except the change in pulse-fraction there is no significant change in the pulse-profiles.

\begin{figure}
\centering
\includegraphics[height=8cm,width=14cm]{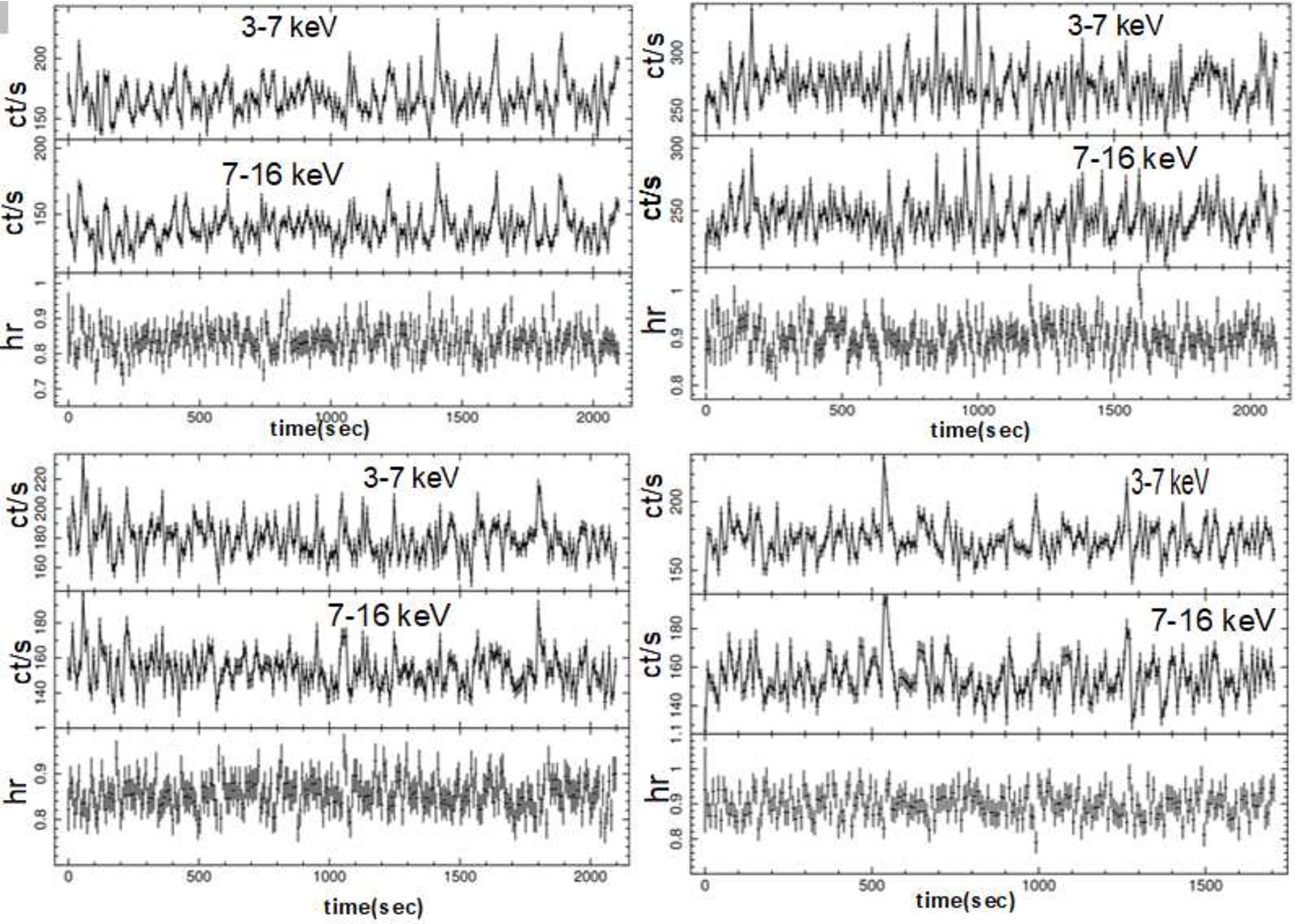}
\caption{\footnotesize Hardness ratio of four observation (a), (b), (c) and (d) obtained by dividing the X-ray photon count rate of 16 keV by 3-7 keV plotted with the time resolution of 8 s}
\end{figure}

$\;\;$The hardness ratio for the four observations were obtained by dividing the 7-16 keV energy X-ray photons count rate with 3-7 keV energy X-ray photons count rate. The average hardness ratio for four lightcurves are $0.9\pm0.03$, $0.897\pm0.035$, $0.8563\pm0.031$ and $0.837\pm0.031$ respectively in fig 5(a), 5(b), 5(c), 5(d). The average hardness ratio for the four observations do not vary significantly. Also there is no noticeable change in hardness ratio in any of the four lightcurves during burst. Therefore the hardness ratio may not be correlated with flares from the source. We can check this invariance in the hardness ratio  by studying the energy spectra of the source.
\begin{figure}
\centering
\includegraphics[height=11cm,width=9cm]{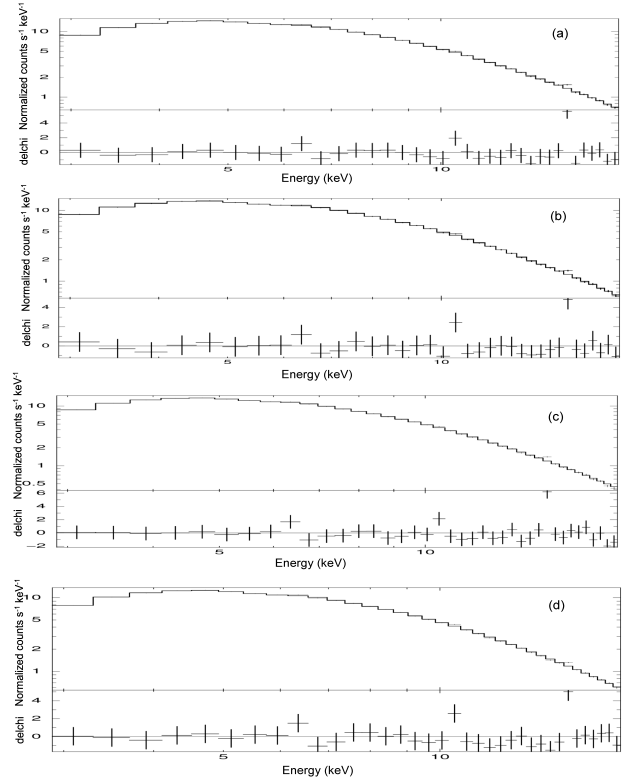}
\caption{\footnotesize Energy spectra of the four observation (a), (b), (c), (d) along with their best-fit spectra. The lower panel of each spectrum is the delchi of the fit, which is the ratio of the difference of the observed data and model value to the corresponding error.}
\end{figure}

$\;\;$The energy spectra for four observations are shown in fig.6 . The models we have used for fitting these spectra are PHABS to estimate the photoelectric absorption of the photon by an intersteller medium, POWERLAW, HIGHECUT for the non-thermal emission of the source and GAUSSIAN for the iron line. The Gaussian peak for an iron line is fixed at  6.7 keV. The  best fit parameters of the fit are shown in Table 1. For all observations the power-law index are nearly same for first three is $\sim$1 which are at fig. 6(a), 6(b), 6(c) where as it is $\backsim$1.1 for fig. 6(d). None of the best fit parameters changed significantly for the four spectra. Considering the distance of SMC X-1 from observing point at 65 kpc (\citealt{Keller+etal+2006, Naik+etal+2004}) the luminosity of source for the four observations are measured which are 6.787$\times$10$^{38}$ ergs$^{-1}$, 6.302$\times$10$^{38}$ ergs$^{-1}$, 6.211$\times$10$^{38}$ ergs$^{-1}$ and 5.862$\times$10$^{38}$ ergs$^{-1}$ respectively for Fig. 6(a), 6(b), 6(c) and 6(d). The variation of the $n_{H}$ value may be due to the partial obscuration of the neutron star by the precessing accretion disk or X-ray eclipses. Another possibility for the variation may be an artifact of simple POWERLAW model or the tails of the soft excess components affecting our result (\citealt{Inam+etal+2010}). The photoelectric absorption by the interstellar matter is dominated in lower energy range  but as we can't go below 3 keV energy due to the limitation of our instrument, its not possible to{ constraint the $n_{H}$ value precisely.
\begin{table}
\begin{center}
\begin{tabular}{|c|c|c|c|c|}
\hline 
observation  & a & b & c & d \\ 
\hline 
n$_{H}$ & 2.062$\pm$ 1.023 & 1.014$\pm$ 1.06 & 1.604$\pm$ 1.043 & 1.968$\pm$ 1.019 \\ 
 
$\alpha$ & 1.066$\pm$0.202 & 1.004$\pm$0.219 & 1.084$\pm$0.218 & 1.122$\pm$0.202\\ 

$E_{fold}$& 17.868$\pm$4.726 &  16.521$\pm$4.457 & 17.723$\pm$5.078 &  17.080$\pm$4.143 \\ 

$E_{cutoff}$ & 5.998$\pm$1.045 &5.870$\pm$1.074 & 5.837$\pm$1.147 & 6.002$\pm$1.218\\ 

flux & 1.345 & 1.247 & 1.229 & 1.160\\

$\chi_{\nu}^{2}$ & 1.682 & 1.375 & 1.890 & 1.563 \\ 
\hline
\end{tabular} 
\end{center}
\caption{\footnotesize The best-fit parameters of the fit. The $\chi_{\nu}^{2}$ is the reduced chi-square of the fit for 29 degrees of freedom. n$_{H}$ is the hydrogen column density of there intervening interstellar matter.$\alpha$ is the power-law index. E$_{cutoff}$ and E$_{fold}$ are the cutoff energy and e-folding energy of the model highecut expressed in keV. The measured flux is for energy range 3-18 keV and in the unit of 10$^{-9}$ ergs$^{-1}$cm$^{-2}$.}
\end{table}
\subsection*{Variation of the spectral parameter with orbital phases and flux}
$\;\;$To study the variation of different spectral parameters over the binary orbital, the available spectra in energy range $3-18$ keV are fitted with the models discussed above, the reduced $\chi^{2}$ of the fitting lies between $1$ to $2$. We plotted the hydrogen column density ($n_{H}$) and photon index with the orbital phases. The spectral flux between the energy range $3-18$ keV were also plotted with the phase. From the left of the figure (7) we can see that there is no abrupt change in the photon index and $n_{H}$ value with orbital phase. It is evident from the figure 7(c) that the flux lies between the $0.992\pm0.001\times10^{-9}$ to $1.7326\pm0.001\times10^{-9}$. The maximum of flux being at an orbital phase of $0.44$ and the minimum at $0.16$. The minimum of $n_{H}$ value is $1.01662\times10^{22}cm^{-2}$ and it happen when the flux is maximum. As evident from the Fig. 7(a),7(c)and 7(d) that the spectrum is bit softer at a phase $0.17$ with the maximum value of the photon index i.e. $1.16604\pm0.3048$ along with the maximum value of hydrogen density column $2.4156\times10^{22}$. The spectrum becomes harder with photon index $0.9574\pm0.02$ as the flux becomes maximum. From figure 7(b) and 7(d) we can see there is no correlation of spectral softening or hardening with phase and flux respectively. The small hardening of the spectrum with the minimum hydrogen density column may be because of the low absorption of the hard X-ray by the intersteller medium which also result in maximum flux observed at that phase. Similarly the small softening of spectrum may be a result of increase in absorption of X-rays by interstellar medium. The overall variation of the flux may be due to variation in the accretion rate of the neutron star.
\begin{figure}
\centering
\includegraphics[scale=.5]{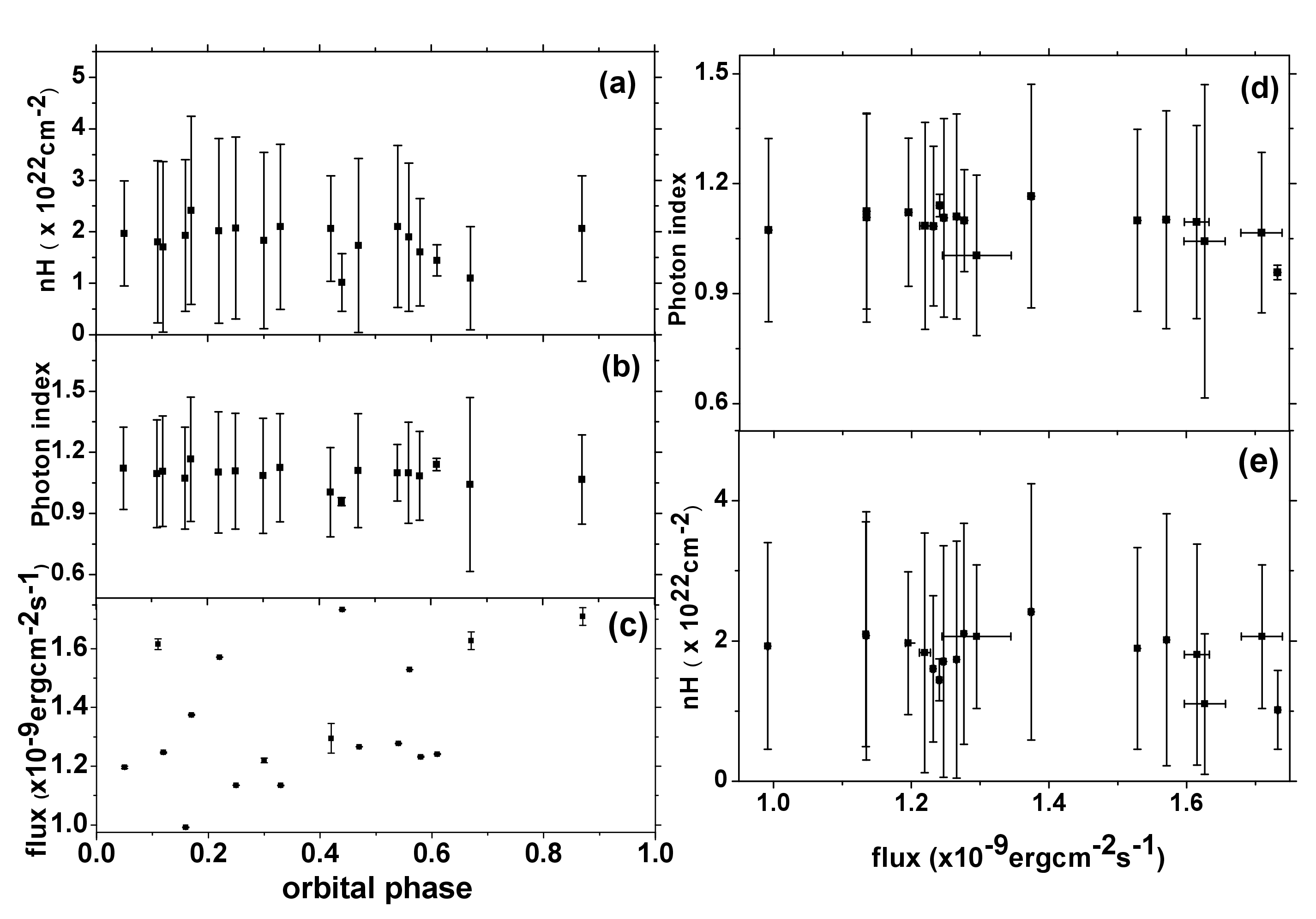}
\caption{\footnotesize The variation of the $n_{H}$, photon index of PL and flux are given by (a),(b) and (c) respectively, where as figures (d) and (e) gives the variation of $n_{H}$ and photon index with the flux respectively.}
\end{figure}

\section{Correlation of the flares with the time, phase peak to peak ratio and orbital phase}
$\;\;$The number of flares observed depends on the total observation time as shown in Fig.8a. Thus it may be concluded that if observation time is longer then more number of flares may be observed. We did not found any correlation of the flare fraction with the rms variability of the lightcurve of the source. The rms variability were found to lie between $\backsim$30-33$\%$ with an average of $\backsim$31$\%$. To find out the correlation of the flares with the pulse profile peak-to-peak ratio we divide the flare fraction of a particular observation by the time so the flare fraction is now measured in $\%$ hr$^{-1}$ and plot with the peak-to-peak ratio of secondary to primary peaks of the pulse profiles. The flare fraction per hour shows some correlation with the peak-to-peak ratio (Fig.8b), there is a relative increase in the flare-fraction per hour. However it is evident from Fig.8c that no correlation of the flare-fraction with the orbital phase exists. The flare-fraction/hr is constant from $\backsim 0.05$ to $\backsim 0.3$ then start decreasing except near 0.2 where a change is observed, it is noted that near 0.56 an increase in the flare fraction is also observed. These variation of flare-fraction with orbital phase may be due to the varying accretion rate of the neutron star which changes the number of bursts observed. To know whether the flaring is the reason behind the change in photon index which results in the softening or hardening of the spectrum we plotted the photon index of the spectrum with respect to the flare-fraction/hr of different observations (Fig.8f). It is observed from the figure(8f) that one cannot correlate the change in photon index with the change in the flare-fraction, however the spectrum looks harder when the flare-fraction/hr($\%$) is $\backsim$10. As there is no correlation of photon index with flare-fraction, one can conclude that the hardening of the spectrum in this case may be due to the decrease in absorption by the intersteller medium not due the increase in the flaring rate.

\begin{figure}
\includegraphics[scale=.5]{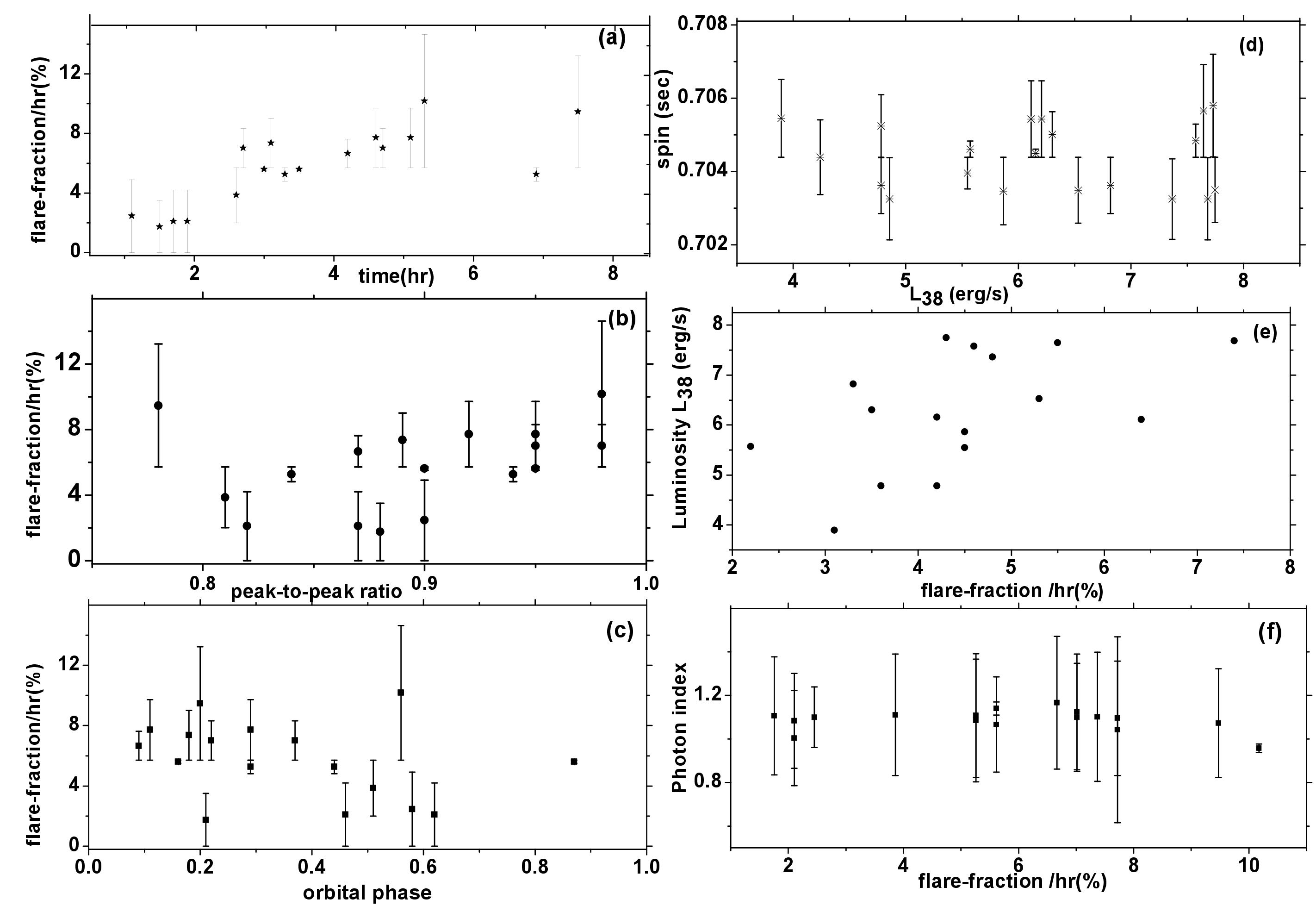}
\centering
\caption{\footnotesize The variation of the flare fraction with the time of the observation (a), with the pulse profile peak-to-peak ratio (b) and with the orbital phase (c). For (b) and (c) the average flare fraction i.e. flare fraction per hour is plotted against peak-to-peak ratio and orbital phase respectively. The variation of the spin with the luminosity is plotted in (d) and in (e) the variation of luminosity with the flare-fraction is shown. Figure (f) gives the variation of photon index of the spectrum with the flare-fraction per hour.}

\end{figure}
The variation of the spin period of the neutron star with the luminosity of the source is given in Fig.8(d). The spin periods were obtained by using a timing tool EFSEARCH in  orbital corrected data obtained by following the method described in Section 4. The spin period in our observations lies between $\backsim$ 0.7033 and $\backsim$ 0.7057 with the luminosity ranging from 3.9$\times$10$^{38}$ to 7.8$\times$10$^{38}$ erg/s. The spin-up rate is found to be directly proportional to $PL_{38}^{3/2}$ where $(P)$ is the spin of the neutron star and $(L_{38})$ is the luminosity of the source, as the spin up rate $\dot{P}$ is small about $\sim$3.279$\times$10$^{-11}$Hzs$^{-1}$ (\citealt{Davison+etal+1977, Wojdowski+etal+1998}) and the two high states is separated by $\backsim$50 days, so there is negligible spin up. Hence the variation of the spin $(P)$ with luminosity $(L_{38})$ is $PL_{38}^{3/2}$=constant (\citealt{Ghosh+etal+1979}). In our case we do not see such variation of spin of the source with 
its luminosity  which may be because of the fact that the data we have considered are from few months observations of the source. It can be possible to find the relation between luminosity and spin and also possibly the relation between spin-up rate, spin and luminosity with the help of long term study of the source. But in the source like SMC X-1 it is harder to study the spin-luminosity relation because of superorbital modulation of luminosity which causes the luminosity to vary in an aperiodic manner. The luminosity of the source increases with the increase in flares as evident from the Fig.8(e), i.e. the luminosity of the source have strong positive correlation with the flares.

\section{Discussion}
\label{sect:discussion}

$\;\;\;$The source SMC X-1 emits type II bursts with the mean recurrence time of $\backsim$800 s and 2.5$\%$ of the observed time the source was bursting. The type II burst are due to the Lightman-Eardley instability which develops in the viscous accretion disk. The average number of burst per hour is found to be $\backsim$4-5. Thus SMC X-1 is a HMXB ``bursting pulsar". We notice a large burst of very short duration and one with long duration with multiple peaks. However the observed pulse profiles do not change its shape despite the fact that the different observation were carried out in different bursting state signalling us that the accretion disk geometry has not changed because of burst. The peak-to-peak ratio of the spin-phase increases with the increase in the flaring rate which may be due to greater rate of accretion of matter in the cooler pole as compared to the hotter pole of the neutron star during bursts             (\citealt{Moon+etal+2003}). We suspect that the increase in the accretion rate above  leads to the above instability, so the increase in the accretion rate may cause nearly equal accretion of matter in both the poles resulting in nearly equal primary and secondary peaks as observed in pulse profile. The luminosity of the source is found to increase with increase in the flaring rate as the conversion of matter into radiation takes place in higher rate than during normal state. 

$\;\;$The Lightman-Eardley(LE) instability is seen when the radiation pressure become dominant and major contributor to the total pressure. The instability is followed by the thermal and surface density instability. The global nature of the instability were investigated by the \citealt{Taam+etal+1984} and \citealt{Lasota+etal+1991} and found that the instability results in bursts which recur after few seconds and were similar to the three outbursts observed in the lightcurve of GRO J1744--28 during 1996 (\citealt{Cannizzo+1996}),  which had outburst of $\backsim$ 10 s and recurrence times of $\backsim$ 1000 s. According to Cannizzo the fast timescale seen by \citealt{Taam+etal+1984} and \citealt{Lasota+etal+1991} were because they set the viscosity parameter '$\alpha$' (\citealt{Shakura+etal+1973}) equal to one and considered the inner radius ($r_{inner}$) of the accretion disk equal to the neutron star radius. ($R\backsim$10km). It is pointed out that for outbursts to occur in GRO J1744--28 the accretion rate must exceed the critical value by a small amount at which the radiation pressure is comparable to the gas pressure. Taking the reasonable value of $\alpha$ (less than 1 and $r_{inner}$ greater than $R$ the viscous timescale at the critical condition have been found to be equal $t_{\nu,crit}$=1200$r^{0.58}_{8}m^{0.79}$ s, where $r_{8}$=$r_{inner}$/10$^{8}$cm, $m=M/M_{\odot}$ and M is the mass of the neutron star which is in good agreement with the recurrence time of the burst in GRO J1744-28, for $r_{inner}$=10$^{7.5}$ cm and $M$=1.4$M_{\odot}$(\citealt{Cannizzo+1996}), $t_{\nu,crit}\backsim$800 s. The burst from SMC X-1 is similar to that of GRO 1744--28 but we note that the post-flare dip is absent in the former. The absence of the dip in the former may be due to slow accumulation of matter after the burst or because of release of only certain amount of accumulated matter during the burst so after the burst the intensity is found in the persistent level.Another possibility for no dip observed in the SMC X-1 after burst may be due increase in accretion rate just after the burst.

$\;\;$There were no correlation of flare-fraction/hr, $n_{H}$ and flux with the orbital phase. There exist no evidence of change in the nature of spectrum because of flares. The small softening or hardening of the spectrum were not due to flares but may be due to varying interstellar absorption. The spectrum of burst for SMC X-1 shows similar properties like GRO J1744--28 as there is no spectral softening and also inconsistent with the blackbody, also the photon index is $\backsim$1.2 and the high cutoff energy of 14 keV (\citealt{Swank+etal+1996}). We may conclude that the bursts in the two X-rays binaries are due to LE instability with the comparable magnetic field. Due to the presence of low magnetic field the transition region (between the radiation dominated and gas pressure dominated region) is located near the inner edge of the disk (\citealt{Li+etal+1997}),therefore the instability develops inside the disk which is not carried far and heats the surrounding matter because of increasing viscosity. The heated matter is then accreted onto the neutron star producing burst. From \citealt{Moon+etal+2003} in the transition region between the radiation dominant and pressure dominant region the viscous time scale is $\alpha=216\dot{M}_{17}/{t_{visc}^{\frac{3}{2}}}$, $\dot{M}_{17}$ is the accretion rate in the order of 10$^{17}$, $t_{visc}$ is the viscous time scale, taking the typical value of $\dot{M}_{17}$=20gms$^{-1}$ and $t_{visc}$=800 s, $\alpha$ $\backsim$0.16. If the recurrence of the burst occur in the same timescale as that of the viscous timescale then we can say the $\alpha$ $\backsim$0.16. The accumulation of huge amount of matter for the short time may lead to the large bursts for a short time, however if accumulation of matter takes slowly resulting in the instability which develops in large area for long time of disk causing long bursts with multiple peaks.

The aperiodic variability of the power spectra can be a possible low frequency quasi-periodic oscillation(QPO) because of interaction of the magnetosphere of the pulsar with the inner edge of the accretion disk as observed in EXO 2030+375 and Cen X-3 or "peaked low frequency noise'' (LFN) seen in LMXRB's Sco X-1 and GX 17+2 (\citealt{Angelini+etal+1991}). Our observation suggests that the flaring activity of the pulsar bears no correlation with this aperiodic variability as seen in figure 2 the aperiodic variability exists whether number of flare increases or decreases. The rms variability of the lightcurves of the source is observed to be independent of the flares implying the source were varying equally all the time.

\normalem
\begin{acknowledgements}
The research has made use of data obtained through the High Energy Astrophysics Science Archive Research Center (HEASARC) Online Service, provided by NASA's Goddard Space Flight center. BCP would like to acknowledge SERB-DST for a research grant EMR/2016/005734. The authors would like to thank the anonymous reviewer for his/her valuable comments and suggestions to present in this form. 
\end{acknowledgements}
  
\bibliographystyle{raa}
\bibliography{bibtex}

\end{document}